\documentclass[aip,jcp,preprint]{revtex4-1}

\usepackage{graphicx}
\usepackage{subfigure}
\usepackage{amsmath}
\usepackage{amssymb}
\usepackage{latexsym}
\usepackage{multirow}
\usepackage{color}
\usepackage{verbatim}
\usepackage[hidelinks]{hyperref}

\newcommand{\ve}[1][K]{\mathbf{#1}}

\graphicspath{{./fig/}}

\newcommand{\etal}{{\it et al. }}
\begin{document}


\title{Effective diffusivity of Brownian particles in a two dimensional square lattice of hard disks}

\author{M. Mangeat}
\affiliation{Univ. Bordeaux, CNRS, Laboratoire Ondes et Mati\`ere d'Aquitaine (LOMA), UMR 5798, F-33405 Talence, France}
\affiliation{Center for Biophysics \& Department for Theoretical Physics, Saarland University, D-66123 Saarbr\"ucken, Germany}
\author{T. Gu\'erin}
\affiliation{Univ. Bordeaux, CNRS, Laboratoire Ondes et Mati\`ere d'Aquitaine (LOMA), UMR 5798, F-33405 Talence, France}
\author{D. S. Dean}
\affiliation{Univ. Bordeaux, CNRS, Laboratoire Ondes et Mati\`ere d'Aquitaine (LOMA), UMR 5798, F-33405 Talence, France}
\affiliation{author to whom correspondence should be addressed: david.dean@u-bordeaux.fr}

\begin{abstract}
We revisit the classic problem of the effective diffusion constant of a Brownian particle in a square lattice of 
reflecting 
impenetrable hard disks. This diffusion constant is also related to the effective conductivity of non-conducting and infinitely conductive disks in the same geometry. We show how a recently derived Green's function for the periodic lattice can be exploited to derive a series expansion of the diffusion constant in terms of the disk's volume fraction $\varphi$. Secondly we propose a variant of the Fick-Jacobs approximation to study the large volume fraction limit. This combination  of analytical results is shown to describe the behavior of the diffusion constant for all volume fractions. 
\end{abstract}


\maketitle

\section{Introduction}

Determining the transport properties of tracer particles in complex media such as colloidal crystals and suspensions, porous media or living cells is an important and complex physical problem \cite{bressloff2013stochastic,burada2009diffusion,burada2008entropic,hofling2013anomalous,dentz2011mixing,dean2007effective}.   In the case of particles diffusing in crowded or confined environments, the presence of obstacles (which can be seen as entropic traps or barriers) hinders their motion, leading to a decreased effective diffusivity~\cite{malgaretti2013entropic,burada2009diffusion,burada2008entropic,jac1967}. The relation between this diffusivity and the geometry of the confinement is known to be non-trivial and has interested physicists over the last few decades~\cite{yang2017hydrodynamic,burada2008entropic,reguera2006entropic,kalinay2006corrections,malgaretti2013entropic,Malgaretti2016}.

\begin{figure}
\begin{center}
  \includegraphics[width=10cm]{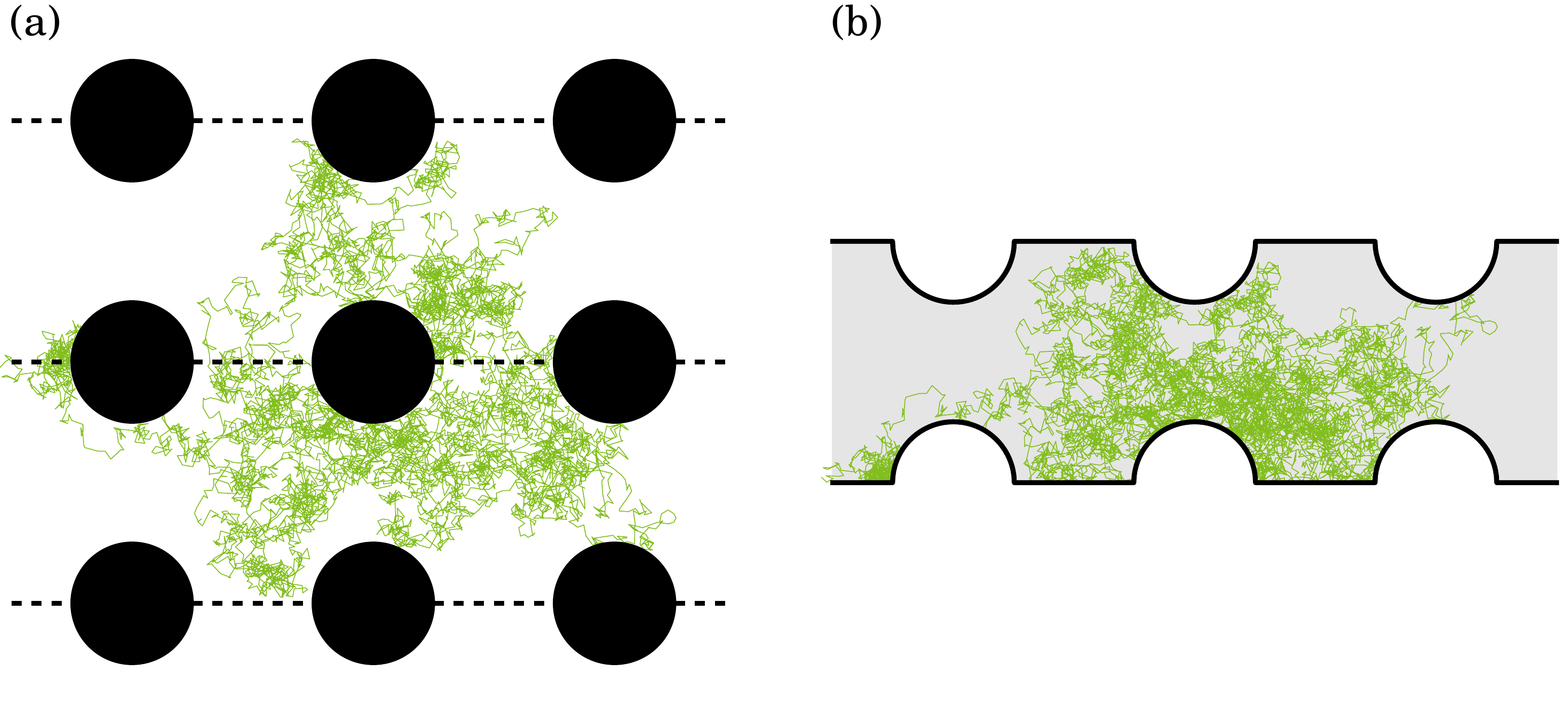}
  \caption{Equivalence of the dispersion of Brownian particles inside an array of obstacles~{\bf (a)} and the dispersion in a periodic channel~{\bf (b)}. From the symmetries of the periodic array of disks, the dashed lines represented in~{\bf (a)} can be replaced by ghost reflecting boundaries which correspond to the boundaries of the channel. The same numerical realization of  Brownian motion is shown in both domains.   \label{figure_dagdug}}
\end{center}
\end{figure}

A standard, and extensively studied  model of dispersion in a complex crowded system is that of a Brownian tracer particle which diffuses in an environment of fixed hard spheres (Fig. \ref{figure_dagdug}). The late time diffusivity $D_e$ of a Brownian tracer  system is related to   the effective conductivity  of a heterogenous medium of fixed non-conducting spheres embedded in a conducting medium (we recall this result below), and results for this problem can be traced back to Maxwell~\cite{maxwell_1881_treatise} and Rayleigh~\cite{lordrayleigh_1892_influence}. For example, in the dilute limit where the volume fraction $\varphi$ of obstacles tends to zero, the calculation of $D_e$ can be formulated as a Laplace equation in infinite space, leading to the well known formula~\cite{maxwell_1881_treatise,jeffrey_1973_conduction, batchelor_1976_brownian} \begin{equation}
\label{DeLD}
\frac{D_e}{D_0} = 1 - \frac{\varphi}{d-1} + {\cal O}(\varphi^2) 
\end{equation}
where $d$ is the spatial dimension,  $D_0$ the microscopic (local) diffusivity and $\varphi$ the volume fraction of obstacles. For higher values of $\varphi$, the above formula is not accurate, and a partially renormalized formula has been proposed  by Kalnin \etal \cite{kalnin_2001_calculations} who obtained
\begin{equation}
\label{DeMaxwell}
\frac{D_e}{D_0}  \simeq \frac{d-1}{d-1+\varphi}.
\end{equation}
The above expression turns out to be a much better approximation for the effective diffusivity in the dilute regime than the leading order expression (\ref{DeLD}). This resummation is equivalent to that proposed by Maxwell for the conductivity problem. However, its mathematical status is rather unclear and only the term first order term in $\varphi$ is strictly exact. It is obtained via an effective medium approach, assuming spherical symmetry around a single obstacle, with an effective diffusivity at infinity determined self-consistently and representing the effect of other obstacles. 
 
In this paper, we show that, for regular square arrays of obstacles, the renormalized expression (\ref{DeMaxwell}) is exact up to (and including) the order $\varphi^4$. This explains why the renormalized formula is accurate up to relatively large volume fractions, around $\varphi\simeq 0.5$ in 2D, for such volume fractions it is rather surprising that a result obtained in the dilute limit should  be relevant.  In fact, our approach uses strong localized perturbation theory~\cite{ward1993} and enables us to obtain exact analytical expressions for the coefficients of $\varphi$ at all orders. We also show that the asymptotic expansion of $D_e(\varphi)$  can be extracted from the literature  as the result of the multipole expansion method, it agrees with our results (derived with a new point of view).  These considerations are the first contribution of the present paper. 

Next, in 2D, we will also consider the opposite limit where $R$ approaches the value $R_c$ (half the square lattice period) at which the obstacles  begin to touch each other, leaving only small openings for the tracer particle to pass to the next pore. In this limit, the diffusivity vanishes as   $\sqrt{\vert R_c-R\vert}$.  
In fact, one can see the 2D problem as equivalent to the problem of a tracer particle in a \textit{channel} \cite{dagdug2012diffusion} (figure \ref{figure_dagdug}b), by adding a ghost reflecting boundary condition identical to the zero flux condition of the boundary of the channel. For channel geometries, we can apply  well known  Fick-Jacobs approximation, which maps the 2D problem onto  the 1D problem of a diffusing particle in an effective, entropic, potential. It is known\cite{dagdug2012diffusion} that this approximation can be used to estimate the effective dispersion in the limit $R\to R_c$, and even improved by introducing effective local diffusion constant $D(x)$ for the effective one dimensional dynamics\cite{berezhkovskii2011time}. The second contribution of this paper is to propose an alternative expression for the local effective diffusivity $D(x)$ in the effective 1D description: 
\begin{align}
D(x)\simeq D_0 \{ 1-[\hat{\ve[e]}_x\cdot \ve[n](x)]^2/3\}, \label{EffD}
\end{align}
where $\ve[n]$ is the normal vector to the boundary and $\hat{\ve[e]}_x$ the unit vector in the direction parallel to the channel. The above formula can be seen as a natural renormalization of the effective diffusivity tensor when studying diffusion between two hypersurfaces of arbitrary dimensions, and is exact at the same order as other expressions of $D(x)$ proposed in the literature~\cite{zwanzig1992diffusion,reguera2001kinetic,kalinay2006corrections}. In the limit of nearly touching obstacles, we find that the use of the local diffusivity (\ref{EffD}) leads to an closed form analytical estimate [Eq.~(\ref{normSph})] of the effective diffusivity which turns out to be more accurate than with the use of other choices of $D(x)$ (and which more over do not have simple closed form expressions).  

Taken together, the analytical formulas presented in this work, in the dilute or crowded regime, provide an accurate description of $D_e$ in 2D for almost all values of the volume fraction. The outline of this paper is as follows: in Section \ref{FormalismSection} we present a general formalism to investigate dispersion in heterogeneous media. Then, we investigate the dilute limit (Section \ref{SectionDilute}) employing an exact perturbative approach. Finally we focus on the limit of nearly touching obstacles (Section \ref{SectionTouchingObstacles2D}) within the context of the improved Fick-Jacobs approximation.

\section{Dispersion in a regular array of spherical obstacles: general formalism}
\label{FormalismSection}
We consider a periodic array of hard spheres with each sphere is taken to be of radius $R$ and at the centre of an (hyper)-cube of length $L$. The spheres are impenetrable and we denote by 
\begin{align}
\Omega=L^d(1-\varphi)
\end{align}
 the volume accessible to the tracer in each hypercube. The overall system is constructed by repeating the underlying cell structure periodically over all space. 
 We consider the motion of small tracer particles inside this array, with a microscopic diffusivity $D_0$ (see figure \ref{figure_dagdug}a) and we define the late time effective diffusivity, say in the $x$ direction, via
\begin{align}
D_e=D_{xx}=\lim_{t\to\infty}\frac{\left\langle [x(t)-x(0)]^2\right\rangle}{2t}, 
\end{align}
where $\langle...\rangle$ denotes ensemble average. It is known that this diffusivity can be obtained by solving a partial differential equation problem for an auxiliary function $f$, see for example the macro-transport theory of Brenner and Edwards~\cite{brenner2013macrotransport}, or Refs.~\cite{guerin2015,Guerin2015Kubo,mangeat2017geometry} for the case of arbitrary (possibly out-of-equilibrium) heterogeneous media. Here we use the notations of Ref.~\cite{mangeat2017geometry} and express the diffusivity as
\begin{equation}
\label{DeKubo}
\frac{D_e}{D_0} = 1  
- \frac{1}{\Omega} \int_{\partial \Omega} dS\ \ve[n]\cdot\ve[e]_x f({\bf x}), 
\end{equation}
 where $\ve[n]$ is the unit normal vector, oriented towards the interior of the obstacles,  $\ve[e]_x$ is the unit vector in the $x$ direction and the integration is taken over the surface of a single obstacle. The auxiliary function $f$ is \textit{periodic} in both $x$ and $y$, is harmonic in the available volume
 \begin{align}
 \nabla^2f =0 \label{Laplacianf}
 \end{align}
 and satisfies the boundary condition
\begin{align}
&{\bf n} \cdot  \nabla f({\bf x})  = {\bf n} \cdot  \ve[e]_x   & (\ve[x] \in \partial\Omega) \label{BCf}.
\end{align}
where $\partial\Omega$ denotes the boundary of the obstacles. Note that $f$ is defined up to an unimportant additive constant. The above set of equations are not in general analytically tractable, in particular the 
periodic boundary conditions for $f$ on $\Omega$ present the main obstacle to analytical
progress. The above set of equations can however  be easily solved numerically using standard partial differential equation solvers, eliminating the need for stochastic simulations of the diffusion process itself.

Here, it is useful to make an explicit connection with an electrostatic problem.  If we set $V=f-x$, we see that $V$ is harmonic and satisfies Neumann boundary conditions $\partial_{n}V=0$ at the  surface of obstacles. Furthermore, the periodicity of $f$ implies that $V(x+L)-V(x)=-L$. This means that $V$ is the electrostatic potential in a medium of non-conducting spheres embedded in a medium of conductivity $\sigma_0=1$ submitted to a constant average electric field $\overline E=1$, where the over-bar denotes uniform spatial average. The effective conductivity is defined by the relation $\sigma_e=\overline{\sigma E}/\overline{E}$, and from this we find 
\begin{align}
\frac{\sigma_e}{\sigma_0}=- \frac{  \int_{\Omega}d\ve[r] \  \partial_x V}{L^d \overline{E}} = \frac{1}{L^d}\left( \Omega-\int_\Omega d\ve[r]  \ \partial_x f\right).
\end{align}
Using the divergence theorem and Eq.~(\ref{DeKubo}) we obtain
\begin{align}
\frac{\sigma_e}{\sigma_0}=\frac{\Omega}{L^d} \frac{D_e}{D_0}=(1-\varphi)\frac{D_e}{D_0}.  \label{LinkElectr}
\end{align}
This means that, \textit{up to the multiplicative factor $(1-\varphi)$}, the effective diffusivity of a periodic medium with reflecting obstacles is exactly the same as the  effective conductivity of a medium containing non-conducting elements. This connection is  well known~\cite{dean2007effective,dean2004perturbation,batchelor1974transport,kalnin_2001_calculations} but the above formula is useful to extract formulas on diffusivity from the literature on electrostatics in heterogeneous media.  

From now on, we focus on the case of two dimensions. 

 \section{Exact asymptotic expansions of the effective diffusivity in the dilute limit $\varphi\to0$ in 2D}
\label{SectionDilute}

We now consider dispersion in the dilute limit, $\varphi\to0$. In this Section, we choose units of length such that $L=1$. As mentioned above, the main obstacle to analytical progress is the periodicity condition on $f$. However, in the dilute limit, it is convenient to look at the behavior of $f$ at the vicinity of the obstacles and to solve the equations by requiring that $f$ vanishes at infinity (meaning that that periodic conditions are approximately satisfied).  
The solution then reads
\begin{align}
f\simeq -  \cos\theta R^2/r  \label{fLeadingOrder},
\end{align}
where $R$ is the radius of the obstacles, and $(r,\theta)$ are polar coordinates when the origin is set at the center of the obstacles. The above expression is the solution of Eqs.~(\ref{Laplacianf},\ref{BCf}) in infinite space, and inserting it into (\ref{DeKubo}) yields 
 \begin{align}
 D_e=1+\frac{R}{\Omega} \int_0^\pi d\theta \cos\theta f(R,\theta)=1-R^2 \int_0^\pi d\theta \cos^2\theta =1-\varphi+o(\varphi) ,
 \end{align}
 where $\varphi=\pi R^2/L^2$ is the volume fraction. This is the Maxwell result (\ref{DeLD}) at first order. To go to next order, we consider the problem as a strong localized perturbation problem (as described in Ref.~\cite{ward1993}), in the spirit of boundary layer theory. The function $f$ is assumed to have components at scale $R$, such that its behavior when $R\to0$ \textit{at fixed} $r/R$ is given by
 \begin{align}
f(\ve[x])=R\left[ \Phi_0(r/R,\theta)+R^2 \Phi_1(r/R,\theta)+R^4\Phi_2(r/R,\theta)+ ...\right]\label{Inner},
\end{align}
 with $\theta$ the angle with respect to $\ve[e]_x$, the usual convention for polar coordinates. The first term of this expansion comes from the comparison with Eq.~(\ref{fLeadingOrder}), which enables us to identify $\Phi_0$ as
 \begin{align}
 \Phi_0(\tilde r,\theta)=-\cos\theta/\tilde r,
 \end{align}
 where $\tilde r=r/R$ is the renormalized radius. The magnitude of the next order terms in the expansion (\ref{Inner}) at this stage is not justified, but it will become clear that this expansion is in fact correct. For the outer solution, obtained by considering the limit $R\to0$ when $r$ is fixed, we assume the following ansatz
\begin{align}
\Phi(\ve[x])=R^2[f_1(x,y)+R^2 f_2(x,y)+R^4 f_3(x,y) + ...] \label{Outer},
\end{align} 
where one has anticipated forthcoming calculations by assuming that the underlying expansion parameter is $R^2$. The leading order term is however determined  from the fact that the outer solution and the inner solution must match in the regime $R\ll r \ll 1$, so that the function $f_1$ must behave 	as
\begin{align}
f_1(r,\theta)\underset{r\to0}{\sim} -  \cos\theta/r,
\end{align}
where the right hand side is identified by considering Eq.~(\ref{fLeadingOrder}). The function $f_1$ must be harmonic and periodic in both $x$ and $y$ with period $1$, and can thus be expressed as 
\begin{align}
f_1(x,y)=-2\pi \ \ve[e]_x\cdot \nabla G(\ve[x] ) \label{Def_f1},
\end{align}
where $G$ is the pseudo-Green's function~\cite{barton1989elements} for the unit-square with periodic boundary conditions, i.e. the solution of
\begin{align}
-\nabla^2 G(\ve[x] ) =\delta(\ve[x])-1
\end{align}
with periodic conditions in $x$ and $y$. It turns out that this pseudo-Green function is known in closed form  \cite{mamode2014fundamental}, and is given by
\begin{align} 
G(x,y) =\frac{1}{4\pi}\ln\left\{e^{-\pi (y^2+x^2)}\vert \theta_1(\pi(x+i y)\vert i)\theta_1(\pi(y+i x)\vert i)\vert\right\}+C ,\label{GExplicit}
\end{align}
where, in the above,  $G$ is defined up to an unimportant additive constant $C$, $i^2=-1$, and $\theta_1$ is the elliptic Jacobi theta function, defined as
\begin{align}
\theta_1(z\vert i)=\theta_1(z,q=e^{-\pi})=2 e^{-\pi/4} \sum_{n=0}^\infty (-1)^n e^{-\pi n(n+1)}\sin[(2n+1)z].
\end{align}
Using Eq.~(\ref{GExplicit}), the behavior of $G$ near the origin reads
\begin{align}
G(\ve[x]) \underset{r\to0}{\sim} \frac{\ln r}{2\pi}- \frac{ r^2}{4 }+C'+  A_4  r^4\cos(4\theta)+\mathcal{O}(r^8\cos(8\theta)),
\end{align}
where $C'$ is another unimportant constant, and the coefficient of the fourth order term reads
\begin{align}
A_4=-\frac{\pi}{16}+\frac{\pi^3 \sum_{n=0}^\infty (-1)^n e^{-\pi (n + 1/2)^2} (2 n + 1)^5 }{240\sum_{n=0}^\infty (-1)^n e^{-\pi (n + 1/2)^2} (2 n + 1)   }  \simeq -0.1253... 
 \end{align} 
Using (\ref{Def_f1}), we see that the behavior of $f_1$ near the origin is
\begin{align}
f_1(\ve[x]) \underset{r\to0}{\sim} -\frac{\cos\theta}{r}+\pi r \cos\theta +C'- 8\pi A_4 r^3\cos(3\theta)+\mathcal{O}(r^7)\label{Behaviorf1}.
\end{align}
We now proceed to calculate the inner solution at the next order $\Phi_1$. The boundary condition for $\Phi_1$ is deduced from the expansion in $R$ of Eq.~(\ref{BCf}), 
\begin{align}
& \partial_{\tilde r}\Phi_1\vert_{\tilde r=1}= 0   \label{BCPhin}
\end{align}
Furthermore, $\Phi_1$ is harmonic and its behavior at infinity must match with the behavior of the outer solution (\ref{Behaviorf1}) near the origin, so that  
\begin{align}
\Phi_1(\tilde{r},\theta)\underset{\tilde{r}\to\infty}{\sim} \pi  \cos\theta \tilde{r}.
\end{align}
The solution for $\Phi_1$ can therefore be written as
\begin{align}
\Phi_1(\tilde{r},\theta)=  \pi  \cos\theta \left(\tilde{r}+\frac{1}{\tilde{r}}\right). \label{SolPHI1}
\end{align}
Inserting this expression into the inner expansion (\ref{Inner}) and the expression (\ref{DeKubo}) for $D_e$ leads to
\begin{align}
D_e=1-\varphi+\varphi^2+\mathcal{O}(\varphi^3)
\end{align}
which means that the renormalized formula for $D_e$ is exact at second order. The solution at next order is obtained as follows: the outer function $f_2$ must diverge as $\pi\cos\theta/r$ for small $r$ to match with the corresponding term in Eq.~(\ref{SolPHI1}), meaning that $f_2= -\pi f_1 $, which admits the small $r$ behavior
\begin{align}
f_2(r,\theta)\underset{r\to0}{\sim}  \pi\frac{\cos\theta}{r}-\pi^2 r \cos\theta +... \label{f2Smallr}
\end{align}
The inner solution $\Phi_2$ is
\begin{align}
\Phi_2(\tilde{r},\theta)=-8\pi A_4 \left(\tilde{r}^3+\frac{1}{\tilde{r}^3}\right)+\pi\cos\theta \left(\tilde{r}+\frac{1}{\tilde{r}}\right)
\end{align} 
which is found by requiring that $\Phi_2$ is harmonic, with Neumann condition at $\tilde{r}=1$ and that the cubic term in $\tilde{r}^3$ matches with the corresponding term in the expression of $f_1$ [Eq.~(\ref{Behaviorf1})]  and that the linear term corresponds to the linear term in the expression of $f_2$ [Eq.~(\ref{f2Smallr})]. Inserting this value into 
Eq.~(\ref{DeKubo}) leads to 
\begin{align}
D_e=1-\varphi+\varphi^2-\varphi^3+\mathcal{O}(\varphi^4)
\end{align}
so that the renormalized expression (\ref{DeMaxwell}) is still valid at this order, even though the non-trivial coefficient $A_4$ is already present in the inner solution $\Phi_2$. 

This calculation can be extended iteratively to higher orders. Since all functions $f_n$ can be constructed with the Green's function $G$, the resulting expression for $D_e$ therefore depends on the coefficients appearing in the expansion of the Green's function. We describe in Appendix  \ref{AppendixDetails2D} how to find $D_e$ to all orders. The main outcome of this calculation is that the renormalized result (\ref{DeMaxwell}) is exact up to (and including) the fourth order in $\varphi$, i.e. 
\begin{equation}
 \frac{D_e}{D_0}  = \frac{1}{1+\varphi}+ \mathcal{O}(\varphi^5).
\end{equation}
and this explains why the above formula is so precise up to seemingly unreasonably high values of $\varphi$ (up to 0.5). The expression of the first correction to the renormalized formula can be expressed in terms of the coefficient $A_4$ appearing in the expression of the Green's function: 
\begin{align}
 &\frac{D_e}{D_0}  =   \frac{1}{1+\varphi} - \frac{384 \ A_4}{\pi^2}  (\varphi^5-\varphi^6)+\mathcal{O}(\varphi^7) \simeq \frac{1}{1+\varphi} -  0.61165    (\varphi^5-\varphi^6)+\mathcal{O}(\varphi^7)
\end{align}
 
Finally,  we also check that the present exact analytical approach is in fact equivalent to the multipole expansion approach. 
Using the above mentioned connection between diffusivity and conductivity,  and an analytical series for the effective conductivity for non-conducting inclusions proposed  by Perrins \etal \cite{perrins1979transport} (correcting a result derived by Rayleigh), we see that the multipole expansion approach for our problem is 
\begin{align}
 \frac{D_e}{D_0}=\frac{1}{1-\varphi}\left( 1-\frac{2\varphi}{1+\varphi-\frac{0.305827\varphi^4}{1-1.402958\varphi^8} - 0.013362\varphi^8}  \right) \label{DePerrins}. 
\end{align}
Direct comparison of the above formula with the expansion obtained within our formalism shows that Eq.~(\ref{DePerrins}) is valid up to the order  $\varphi^{16}$.  
 
 \section{The limit of nearly touching obstacles in 2D}  
 \label{SectionTouchingObstacles2D}
We now consider the limit where the space between obstacles vanishes ($R\to L/2$). For the 2D problem, the effective diffusivity vanishes in this limit, $D_e\to0$. As remarked in Ref.~\cite{dagdug2012diffusion}, and mentioned above, 
adding a ghost boundary condition identical to the zero flux condition at the boundary of an effective channel does not change the effective diffusivity in the $x$ direction,
which can thus  be obtained by considering Brownian motion inside a channel of local height $h(x)$ 
\begin{equation}
\label{hSph}
h(x) = \frac{L}{2}  - \sqrt{R^2-x^2} \ \Theta(R-|x|),
\end{equation}
where $\Theta$ corresponds to the Heaviside function. The above expression is valid for $-L/2<x<L/2$ and is then repeated with period $L$. 

When $R\to R_c=L/2$, the effective diffusivity becomes limited by the crossing of the narrow regions between the obstacles, separated by a length $a$ at the neck.  In these regions, the typical channel height $h$ is small compared to the longitudinal dimensions $L$. In this limit, it is thus appropriate to consider an adiabatic type of  approximation, where one assumes a much smaller relaxation time in the lateral direction than in the longitudinal one. This approximation  is known as the Fick-Jacobs approximation, and, upon its application, the problem is mapped onto a one dimensional dynamics for the coordinate $x(t)$. The effects of the geometry are incorporated via an effective potential 
\begin{equation}
\Phi(x)=-k_BT\ln h(x), \label{ph}
\end{equation}
acting on the particle at the lowest order, and by an effective one dimensional diffusivity $D(x)$ which is modified at higher orders. The resulting effective one dimensional diffusion equation, for the marginal probability distribution along the channel, takes the form
\begin{equation}
\partial_t p^*(x,t) = \partial_x \{D(x)[\partial_x p^*(x,t)+\beta p^*(x,t)\partial_x\Phi(x)]\}.
\end{equation}
with $\beta=1/k_BT$. We notice that the potential must always take the form given in Eq.~(\ref{ph}) as for a finite system the marginal equilibrium probability distribution for the variable $x$ is given by
\begin{equation} 
p^*_{ \mathrm{eq}}(x) = \frac{h(x)}{\int_0^L h(x')dx'}
\end{equation}
as the, joint, full two dimensional equilibrium distribution is uniform.

The effective diffusivity for the resulting one dimensional diffusion equation can be computed using the  exact   Lifson Jackson formula~\cite{lifson1962self}
\begin{equation}
\label{LJformula}
D_e = \frac{1}{\langle h(x) \rangle \langle [D(x) h(x)]^{-1} \rangle},
\end{equation}
where here $\langle ... \rangle$ denotes the uniform average over one period. 

At leading order in the limit $h/L\to0$, the height varies smoothly and the effective local diffusivity is $D(x)\simeq D_0$. This approximation is called the \textit{basic} Fick-Jacobs  (FJ) approximation. In this case, the averages in Eq.~(\ref{LJformula}) have been performed analytically by Dagdug \etal \cite{dagdug2012diffusion}, leading to 
\begin{equation}
\label{FJSph}
\frac{D_e}{D_0} \simeq \left[(1-\varphi)\left(1-\lambda- \frac{\pi}{2} + \frac{2}{\sqrt{1-\lambda^2} }\arctan \sqrt{\frac{1+\lambda}{1-\lambda}}\right)\right]^{-1},
\end{equation}
where the parameter
\begin{align}
\lambda=2R/L
\end{align}
becomes unity when the obstacles touch. This expression recovers the asymptotic result for $\lambda \rightarrow 1$, first obtained by Keller \cite{keller1963conductivity},
\begin{equation}
\frac{D_e}{D_0} \simeq \frac{\sqrt{2(1-\lambda)}}{\pi(1-\varphi)}.  \label{LeadongOrderDeff}
\end{equation}

The next order correction to the Fick-Jacobs approximation (\ref{FJSph}) in terms of a one-dimensional description leads to an effective local diffusivity depending on the local height, the parameter  $\varepsilon=h'$ is assumed to be small and the correction to (\ref{FJSph}) to leading order leads to 
\begin{align}
D(x)=D_0\left[1-\frac{h'(x)^2}{3}\right] \label{EffectiveLocalD}
\end{align}    
where the local slope of the channel is considered as a small parameter. The above formula is exact at order $\varepsilon^2= h'^2$ (included). However for the effective height profile generated by a square array of disks we clearly see that $h'(x)$ diverges for $x=\pm R$.
(Note that when the height profile is actually discontinuous, the standard perturbative approaches to computing the diffusion constant beyond the FJ approximation have to be modified \cite{mangeat2018dispersion}.)

Several partially resummed formulas have been proposed for $D(x)$ that are compatible with the above expression, 
\begin{align}
D(x)=
\begin{cases}
D_0/[1+h'(x)^2/3] & (\text{Zw})\\
D_0/[1+h'(x)^2]^{1/3} & (\text{RR}) \\
D_0\ \text{arctan}(h'(x) ) /h'(x) & (\text{KP})
\end{cases} \label{OtherDx}
 \end{align}
which were proposed by Zwanzig (Zw) \cite{zwanzig1992diffusion}, Reguera and Rub\'i (RR) \cite{reguera2001kinetic} and Kalinay and Percus (KP) \cite{kalinay2006corrections}, respectively. 

Here we propose an alternative form for $D(x)$, based on the argument that the generalization of Eq.~(\ref{EffectiveLocalD}) for the diffusion between two hyper-surfaces of location $y=\pm h(x_1,...,x_{n-1})$, where $n$ is the spatial dimension. The argument is given in the Appendix \ref{appd1}. In this case we find that the effective local diffusion tensor is given by  
\begin{align}
D_{ij}(x)\simeq D_0\left[1-\frac{1}{3} (\partial_{x_i}h) (\partial_{x_j}h) \right]\label{d1}.
\end{align}
We note that the approximate formulas given in Eq. (\ref{OtherDx}) have the peculiar properties that the effective diffusion constant tends to zero where $h'(x)$ diverges - that is to say if the 
channel profile has a kink. This is clearly unphysical as only a drastic narrowing of the channel can reduce the diffusion constant. To remedy this problem we recall that the local normal
of the surface has components
\begin{equation}
n_i(\ve[x])= \frac{\partial_{x_i}h}{(1+[\nabla h]^2)^{\frac{1}{2}}}
\end{equation}
 and so to order $\varepsilon^2$,  Eq. (\ref{d1}) can be written as 
\begin{align}
D_{ij}(x)\simeq D_0\left(1-\frac{1}{3} n_i(\ve[x])n_j(\ve[x])\right)
\end{align}
where $ n_i(x)$ is the $i^{ \text{th}}$ coordinate of the normal to the surface. Coming back to our 2D problem, this leads us to propose an alternative form for the effective local diffusion constant $D(x)$ reads
\begin{align}
D(x)\simeq D_0\{1-[n_x(x)]^2/3 \} \label{OurChoiceLocalD}
\end{align}
 where $n_x$ is the $x-$ component of the normal to the obstacle, 
 \begin{align}
 n_x(x)=-\frac{h'(x)}{ \{ 1+[h'(x)]^2\}^{1/2} }
 \end{align}
 An additional reason to postulate that the use of a local diffusion constant  (\ref{OurChoiceLocalD}) may lead to more accurate results than its non-renormalized form (\ref{EffectiveLocalD})  is that \textit{ the normal to the surface is the natural relevant quantity that appears in the general equations} (\ref{DeKubo}), (\ref{BCf}) for the dispersion  (whereas the local quantity  $h'(x)$ does not appear). Note that with Eq.~(\ref{OurChoiceLocalD}), the local diffusivity is always finite and positive. 
  
Since $n_x(x) = - x/R$ for the disk, we obtain the, everywhere finite, expression 
\begin{equation}
\frac{D(x)}{D_0} \simeq  1 - \frac{1}{3} n_x(x)^2 = 1 - \frac{x^2}{3R^2}  \Theta(R-|x|).
\end{equation}
Using  Eq.~(\ref{LJformula}) with this local diffusion constant we find the, fully analytical, result
\begin{equation}
\label{normSph}
\frac{D_e}{D_0} \simeq \left[(1-\varphi)\left(1-\lambda- \frac{\sqrt 3 \lambda}{2} \frac{\sqrt 2 \pi \lambda - \ln (2+\sqrt 3)}{2 \lambda^2+1} + \frac{6\lambda^2}{(2\lambda^2+1)\sqrt{1-\lambda^2} }\arctan \sqrt{\frac{1+\lambda}{1-\lambda}}\right)\right]^{-1}. 
\end{equation}
The above expression is the main result of this Section. 

\begin{figure}[t]
\begin{center}
  \includegraphics[width=16cm]{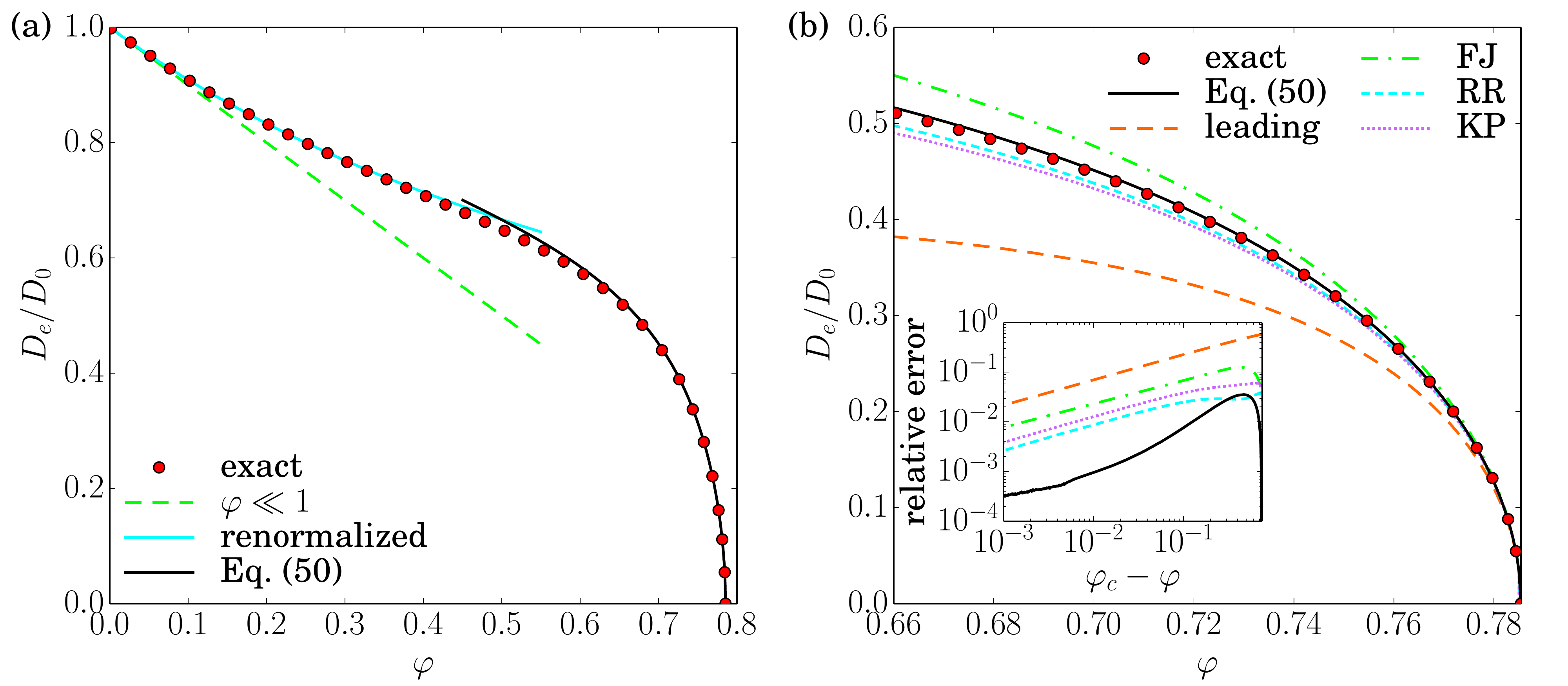}
  \caption{Effective diffusivity of Brownian particles in two-dimensional periodic array of non-attractive obstacles of volume fraction $\varphi$. The circles represent the numerical solution obtained from Eq.~(\ref{DeKubo}-\ref{BCf} ) with a partial differential equation solver (FreeFem++). {\bf (a)} Dashed line: leading order expression (\ref{DeLD}) in the dilute regime. Continuous line: renormalized expression (\ref{DeMaxwell}). The renormalized result is surprisingly valid up to a volume fraction of $\varphi \sim 0.4$. {\bf (b)}~In the crowded limit, the Fick-Jacobs' result (\ref{FJSph}) is represented with dashed lines while other results of the literature shown by Reguera and Rub\'i \cite{reguera2001kinetic} (RR)  and Kalinay and Percus (KP) \cite{kalinay2006corrections} are plotted respectively using dashed-dotted and dotted lines. Our result, valid only for the periodic array of obstacles, Eq.~(\ref{normSph}) is shown using a solid line. The effective diffusivity is then well-approximated for a volume fraction $\varphi \gtrsim 0.68$ while existing results in the literature are only accurate down to $\varphi \gtrsim 0.77$.   \label{figure_circle}}
\end{center}
\end{figure} 
We now turn to the numerical resolution of Eqs. (\ref{DeKubo}-\ref{BCf}) to determine the accuracy of this new approximate formula, and to compare it with the basic Fick-Jacobs approximation shown in Eq.~(\ref{FJSph}), and the improvements proposed in Eqs.~(\ref{OtherDx}).

We expect that the basic Fick-Jacobs approximation should work for narrow channels and so in the region near $\varphi=\pi/4\simeq 0.785$ where the obstacles touch each other and above which the diffusion constant vanishes.
In Fig.\ref{figure_circle}b we see that Eq.~(\ref{normSph}) is valid up to $1\%$ precision all the way down to  $\varphi \approx 0.68$, whereas the basic FJ approximation is valid only down to $\varphi \approx 0.77$ (for the same precision).

Note also that Eq.~(\ref{normSph}) is more accurate than the leading order asymptotics  (\ref{LeadongOrderDeff}). Further more, it is also more accurate than all of the modifications proposed in Eq. (\ref{OtherDx}) for the local diffusivity (and for which we recall that the averages cannot be performed analytically). 

We thus see that the combination  of the results  (\ref{DeMaxwell}) and (\ref{normSph})  perfectly describe the  effective diffusivity in, respectively, the dilute and the crowded limits in 2D.

\section{Conclusion}

We have revisited the old but important problem of the effective diffusivity of a Brownian particle diffusing on a square lattice of hard reflecting disks. This problem is also equivalent to computing the effective conductivity of a medium of nonconducting spheres embedded in a 
conducting background as well as a number of other problems such as flow in porous media and effective dielectric properties \cite{dean2007effective}. Using a recently derived Green's function for the periodic square lattice \cite{mamode2014fundamental}, a perturbative expansion in the volume fraction of the disks was formulated. This formalism allows a systematic almost algorithmic determination of the coefficients of the expansion. We confirm that the first four terms in this expansion have a particularly simple form which explains the the accuracy of Maxwell's effective medium approximation, which is in principle only correct to first order but actually turns out to capture all terms up to fourth order. 

The limit of high volume fraction was treated by mapping the problem onto that of diffusion in a 
one dimensional periodic channel as proposed in \cite{dagdug2012diffusion}. Here we
proposed a new variant of the second order Fick-Jacobs approximation where the height derivative terms in the effective diffusivity are replaced, to the same order, by the expression for the surface normal. This approximation, in common with a number of other suggestions in the literature \cite{zwanzig1992diffusion,reguera2001kinetic,kalinay2006corrections}, avoids the appearance of a negative effective local diffusivity. However it also eliminates the possibility that the effective diffusivity vanishes. Furthermore, comparison with
exact numerical calculations of the effective diffusivity shows that it describes the behavior of the diffusivity with better accuracy and down to much lower volume fractions that existing approximation schemes.

A number of extensions of this work remain open, notably the treatment of different lattices
for hard disks in two dimensions,  in both the dilute and Fick-Jacobs limits. It would be particularly interesting to examine the case of random arrangements of hard disks. 
In addition, extensions to the three dimensional systems with spherical obstacles clearly merits further exploration in both limits.

\section{Acknowledgements}

This work was partially performed with financial support from the Deutsche Forschungsgemeinschaft (DFG) within the Collaborative Research Center SFB 1027. The data that support the findings of this study are available from the corresponding author upon reasonable request.

\appendix

\section{Calculation of $D_e$ at all orders}
 \label{AppendixDetails2D}
Here we describe how to calculate $D_e$ at all orders in the density of obstacles. The notation used is the same as  Section \ref{SectionDilute}. We write the pseudo-Green's function $G$ as\cite{mamode2014fundamental}
\begin{align}
G(x,y)=-r^2/4+\frac{1}{8\pi}\ln[\psi(z)\psi(i\overline{z})\psi(\overline{z})\psi(-i z)]
\end{align}
 with 
 \begin{align}
  \psi(z)=\theta_1(\pi z,e^{-\pi}) = 2 e^{-\pi/4} \sum_{n=0}^\infty (-1)^n e^{-\pi n(n+1)}\sin[\pi (2n+1)z] 
\end{align} 
We can define coefficients $a_p$ which appear in the expansion of $\psi(z)$ near the origin
\begin{align}
&\psi(z)=\sum_{p=0}^\infty a_p z^p,  \\
& a_{2m}=0, \hspace{2cm} 
a_{2m-1}  = \frac{2 (-1)^{m+1} \pi^{2m-1}}{(2m-1)!} e^{-\pi/4}\sum_{n=0}^\infty(-1)^n e^{-\pi n(n+1)}(2n+1)^{2m-1} 
 \end{align}
Our goal is to set up an iterative procedure to calculate the expansion of $D_e$ in terms of the coefficients $a_m$, which can be straightforwardly calculated with the series above.  

We note that all functions $\Phi_k$ (with $k\ge1$) appearing in the inner expansion are harmonic and have Neumann boundary conditions, so that general expression is 
\begin{align}
\Phi_k(\tilde{r},\theta)=\sum_{q=1}^k \gamma_{k,q} \cos((2q-1)\theta)\left(\frac{1}{r^{2q-1}}+r^{2q-1}\right)
\end{align}
where the maximal number of terms $k$ follows from the remark that one additional divergence at most can appear at each iteration. 

Next, we construct a set of functions $g_n$ which satisfy the following requirements: (i) $g_n$ is harmonic on the unit square with periodic boundary conditions, and (ii) the behavior of $g_n$ near the origin is
\begin{align}
g_n(r,\theta)\underset{r\to0}{\sim}\frac{\cos((2n-1)\theta)}{r^{2n-1}}+\sum_{k=1}^\infty\alpha_{n,k}\cos((2k-1)\theta)r^{2k-1}
\end{align}
This means that $g_n$ contains only one singularity of order $1/r^{2n-1}$ near the origin. Next, we remark that such functions $g_n$ can be constructed by using the pseudo-Green's function for the unit square with periodic conditions by using
\begin{align}
&g_1=2\pi \ve[e]_x\cdot\nabla G \label{Defg1}&  \\
&g_{n+1}= \frac{1}{(2n-1)(2n)}(\ve[e]_x\cdot\nabla)^2 g_{n} & (n\ge1)
\end{align}
and this recurrence relation implies that
\begin{align}
\alpha_{n+1,k}=\frac{(k+2)(k+1)}{(2n-1)(2n)}\alpha_{n,k+2}.
\end{align}
The coefficients $\alpha_{1,l}$ can then be determined from the expansion near the origin of $G(\ve[x])$ by using Eq.~(\ref{Defg1}). 

We now express the outer solution as a sum of such functions,  
\begin{align}
f_n(r,\theta)=\sum_{q\ge1}A_{n,q}g_q(r,\theta)
\end{align}
Note that the coefficient of $g_1$ must match the $1/r$ singularity in $\Phi_{n-1}$, the coefficient of $g_2$ must match the $1/r^3$ singularity of $\Phi_{n-2}$, and we repeat this process until we find the last function $\Phi_{n-k}$ that does not contain a $1/r^{2k+1}$ singularity, i.e. the maximal value of $k$ is the integer part of $n/2$. We thus write
\begin{align}
&A_{1,1}=-1 & (n =1),\\
&A_{n,q}=\gamma_{n-q,q} & (n\ge2, q\le\text{floor}(n/2)),  
\end{align}
where the particular case $n=1$ follows from the identification of $\Phi_1$. 

Next, the coefficients of the inner solutions are determined by the fact that  the behavior of $\Phi_n$ must match the linear term $r$ in $f_n$, the cubic term $r^3$ in $f_{n-1}$, and more generally the term $r^{2k-1}$ in $f_{n-k+1}$. Since $f_{n-k+1}$ is a sum over the functions $g_q$, we obtain
\begin{align}
&\gamma_{n,k}=\sum_{q=1}^{q_m(n,k)}A_{n-k+1,q}\alpha_{q,k}\\
&q_m(n,k)\equiv \text{max}[1,\text{floor}(n-k+1)/2 ]
\end{align} 
These formulas enable us to evaluate $D_e(\varphi)$ in terms of a series whose coefficients are calculated iteratively. The first terms are 
\begin{align}
D_e=   1-&\varphi+\varphi^2 -\varphi^3+\varphi^4 -1.611(\varphi^5-\varphi^6) -2.223  (\varphi^7-\varphi^8) -3.049\ \varphi^9 + 
  3.236\  \varphi^{10}  \nonumber\\ 
 & -4.248\ \varphi^{11}
   -4.622\ \varphi^{12} -6.754\ \varphi^{13} + 
  7.446\ \varphi^{14}  -9.953\  \varphi^{15}+
  11.077 \ \varphi^{16}\nonumber\\
  & -15.201\  \varphi^{17}+17.465 \ \varphi^{18} + \mathcal{O}(\varphi^{19}).
\end{align}

\section{Derivation of Equation (\ref{d1})}
 \label{appd1}
Here we derive the approximate relation Eq.~(\ref{d1}) which represents the first correction to the basic Fick-Jacobs approximation. We  consider the stochastic trajectories of a particle $(\ve[x](t),z(t))$ between the surface  of height $z=h(\ve[x])$ and the surface $z=0$. Here $z$ denotes the direction perpendicular to the channel. We start by recalling an exact result for the marginal probability density function $p^*({\bf x},t)$ 
\begin{equation}
p^*({\bf x},t) = \int_0^{h(\bf x)} dz \ p({\bf x},z,t)
\end{equation}
for the position ${\bf x}$ along the channel.   In Ref.~\cite{mangeat2017dispersion} it was shown that 
\begin{equation}
\partial_t p({\bf x},t) = D_0\nabla\cdot\left( \nabla p^*({\bf x}, t) - p({\bf x},h({\bf x}),t) \nabla h({\bf x})\right),\label{mas}
\end{equation}
where $\nabla$ above indicates the gradient operator parallel to the channel, that is to say on the coordinates ${\bf x}$. However, the above is not a closed equation for $p^*({\bf x},t)$ due to the presence of the surface term $p({\bf x},h({\bf x})$ in the advection term. 

We now proceed with a perturbative expansion for the full probability density function $p({\bf x}, z)$ based on the fact that $h({\bf x})$, and thus $z< h({\bf x})$, is small. One writes,
\begin{equation}
p({\bf x},z,t) = p_0({\bf x},t) + z^2 p_2({\bf x},t) + {\cal O}(z^4), \label{expansion}.
\end{equation}
the choice of the expansion in $z^2$ being justified by the no-flux boundary condition at the flat lower surface $z=0$ $\partial_z p({\bf x},z,t)|_{z=0}=0$. The no-flux boundary condition at the upper surface can be written as
\begin{equation}
\left[\nabla p({\bf x},z,t)\cdot\nabla h({\bf x}) -\partial_z p({\bf x},z)\right]_{z=h({\bf x})}=0.
\end{equation}
Inserting Eq.~(\ref{expansion}) into the above,  we find that
\begin{equation}
p_2({\bf x},t) = \frac{1}{2h({\bf x})}\nabla p_0({\bf x},t)\cdot\nabla h({\bf x}).\label{p2}
\end{equation}
Integrating Eq.~(\ref{expansion}) to derive the marginal probability distribution gives
\begin{equation}
p^*({\bf x},t) = h({\bf x}) p_0({\bf x},t) + \frac{h^3({\bf x})}{3}p_2({\bf x},t) + {\cal O}(h^5)  ,
\end{equation}
and using Eq.~(\ref{p2}) then gives to leading order
\begin{equation}
p^*({\bf x},t) = h({\bf x}) p_0({\bf x},t) + \frac{h^2({\bf x})}{6} \nabla p_0({\bf x},t)\cdot\nabla h({\bf x}).\label{pstar}
\end{equation}
The first order solution to the above is given by
\begin{equation}
p_0({\bf x},t)= \frac{p^*({\bf x},t)}{h({\bf x})},
\end{equation}
and using this one recovers the basic Fick-Jacobs approximation. Solving Eq. (\ref{pstar}) to next order by iteration gives
\begin{equation}
p_0({\bf x},t) = \frac{p^*({\bf x},t)}{h({\bf x})} - \frac{h({\bf x})}{6} \nabla \frac{p^*({\bf x},t)}{h({\bf x})}\cdot\nabla h({\bf x}). \label{p0}
\end{equation} 
Now using Eqs.~(\ref{expansion}),(\ref{p2}),(\ref{p0}), we find that the surface term is, to leading order, given by
\begin{eqnarray}
p({\bf x},h({\bf x}),t) = \frac{p^*({\bf x},t)}{h({\bf x})}+ \frac{h({\bf x})}{3} \nabla \frac{p^*({\bf x},t)}{h({\bf x})}\cdot\nabla h({\bf x}).
\end{eqnarray}
Substituting this into Eq.~(\ref{mas}) then leads to the closed form equation for $p^*(\ve[x],t)$:
\begin{equation}
\partial_t p^*({\bf x},t) = \partial_{x_i}\left[D_{ij}({\bf x})\left(\partial_{x_j}p^*({\bf x},t)  -p^*({\bf x},t)\partial_{x_j}\ln(h({\bf x}))\right)\right],
\end{equation}
where $D_{ij}({\bf x})$ is given by Eq. (\ref{d1}).



\end{document}